\documentclass[%
 reprint,
 amsmath,amssymb,
 aps,
]{revtex4-2}
\usepackage{graphicx,amsthm,amsmath,amsfonts,amssymb,xcolor,soul,epstopdf}
\usepackage{dcolumn}
\usepackage{bm}

\begin{document}
\epstopdfsetup{outdir=./}

\title{Circle radius distributions determine random close packing density}

\author{David J. Meer$^1$}
\email{dmeer@emory.edu}
 \author{Isabela Galoustian$^1$}%
 \author{Julio Gabriel de Falco Manuel$^2$}%
\author{Eric R. Weeks$^1$}%
\affiliation{$^1$Department of Physics, Emory University\\ $^2$Physics and Chemistry Institute - IFQ, Federal University of Itajub{\'a} - UNIFEI, Itajub{\'a}, MG 37.500-903, Brazil}

\date{\today}

\begin{abstract}
Circles of a single size can pack together densely in a hexagonal lattice, but adding in size variety disrupts the order of those packings.  We conduct simulations which generate dense random packings of circles with specified size distributions, and measure the area fraction in each case.  While the size distributions can be arbitrary, we find that for a wide range of size distributions the random close packing area fraction $\phi_{\rm rcp}$ is determined to high accuracy by the polydispersity and skewness of the size distribution.  At low skewness, all packings tend to a minimum packing fraction $\phi_0 \approx 0.840$ independent of polydispersity.  In the limit of high skewness, $\phi_{\rm rcp}$ becomes independent of skewness, asymptoting to a polydispersity-dependent limit. We show how these results can be predicted from the behavior of simple, bidisperse or bi-Gaussian circle size distributions.
\end{abstract}

\maketitle

\section{Introduction}

\begin{figure}
    \centering
    \includegraphics[width=0.46\textwidth]{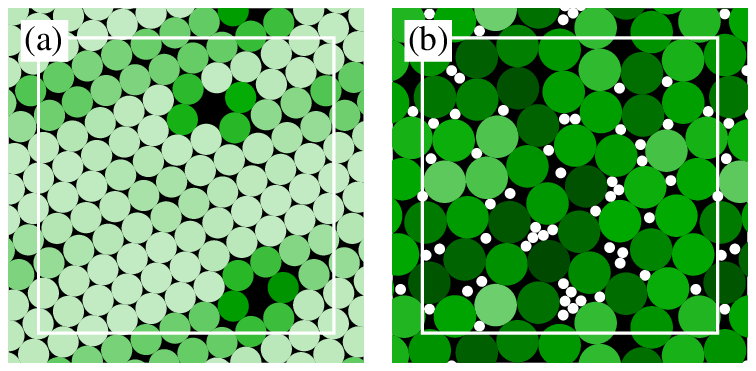}
    \caption{(a) 100 monodisperse circles compressed into a hexagonally ordered configuration, $\phi\approx0.927$. (b) A bidisperse mixture ($\delta=0.6,S=0$) where the smaller circles disrupt the order and create a random packing with $\phi=0.840$. The green color indicates $\psi_6$, a measure of how hexagonally ordered the large circles are relative to each other, and increases towards white. Small circles in (b) are ignored in the calculation of $\psi_6$ and are displayed as white. The thin white square indicates the extent of the periodic boundary conditions.}
    \label{disrupt}
\end{figure}

A classic question is to try to guess the number of marbles in a jar.  The mathematical version of that question is to ask what is the volume fraction occupied by spheres in an amorphous configuration, often termed ``random close packing.''  Our interest is in the two-dimensional version of this question, the random close packing of circles.  For the packing of circles of identical sizes, they in general do {\it not} pack randomly but rather into hexagonal regions, such as Fig.~\ref{disrupt}(a).  To form a ``random close packing,'' one needs a mixture of particle sizes, such as shown in Fig.~\ref{disrupt}(b).  The focus of our work is to understand how the area fraction of such a packing depends on the details of the distribution of particle sizes. The area fraction $\phi$ is the space occupied by the circles divided by the total space the system occupies.  For a system of identical circles packed in a perfect hexagonal lattice, the area fraction is $\frac{\pi\sqrt{3}}{6}\approx 0.907$.  Mixing in other circle sizes can disrupt hexagonal ordering and result in lower values for $\phi$; it is also plausible that mixing in small particles that fit between big particles could increase $\phi$ even in the absence of hexagonal ordering.

Non-random packings of circles have been studied frequently.  For example, Apollonian packings are disordered and pack to high area fractions, but require circles with specific sizes and positions to fill space efficiently \cite{Stephenson,Fernique}.  One can also construct ``Dionysian'' packings that are mechanically stable but at arbitrarily low area fractions \cite{dennis22}.  Other algorithms generate circle packings according to pre-determined networks and graphs \cite{Collins}.  Our interest is in random close packing, although that term is now seen as imprecise \cite{Torquato,atkinson14}. The field is just as rich, and has many applications in nature and industry \cite{Torquato2,Castillo,Danaei,McGeary,Truskett}. In this context the idea of randomness is that circle positions are initially chosen in an uncorrelated fashion, like an ideal gas, and then allowed to move to result in a dense packing of circles such that none are overlapped.

In general, even algorithms that start with random initial conditions usually converge onto hexagonal packings such as Fig.~\ref{disrupt}(a), so long as the circles are all the same size \cite{atkinson14}.  To disrupt hexagonal ordering we turn to packings of circles with a probability distribution of radii $P(r)$.  Distributions of sizes are typically characterized by the polydispersity $\delta$, which can be defined through the first and second moment of the distribution:
\begin{eqnarray}
    \langle r \rangle &=& \int_0^{\infty}P(r) r dr\\
    \langle \Delta r^n \rangle &=& \int_0^{\infty}P(r) (r - \langle r \rangle)^n dr\\
    \delta &\equiv& \sqrt{\langle \Delta r^2 \rangle} / \langle r \rangle.
\end{eqnarray}
Note that the circles and voids between circles both scale with $r^2$, so the area fraction does not depend on $\langle r \rangle$.

In three dimensions, it is well known that increasing $\delta$ increases the packing volume fraction \cite{aim67,sohn68,dexter72,visscher72}, see also references in Ref.~\cite{Desmond2014}.  One would this is also true in two dimensions (2D), in that smaller particles can fit into voids between the larger particles, as shown in Fig.~\ref{disrupt}(b), thus increasing $\phi$.

Of course, $P(r)$ can be nearly any normalized function; the only requirement is that $P(r)$ be  nonzero only for $r>0$.  Knowing polydispersity $\delta$ only gives information about the ratio of the first two moments of the distribution, with all other moments unconstrained.  It is conceivable that there could be many different $P(r)$ with the same $\delta$ but much different achievable random close packing area fractions $\phi$.  In 3D, a perhaps surprising result was found by Desmond and Weeks \cite{Desmond2014}:  knowing just the polydispersity and skewness of a distribution was sufficient to determine $\phi_{\rm 3D}$ of a random close packed sample to within $\pm 0.002$. The skewness is defined as:
\begin{equation}
    S = \frac{\langle \Delta r^3 \rangle}{\langle \Delta r^2 \rangle^{3/2}}.
\end{equation}
Skewness describes the asymmetry of the distribution.  Large positive skewness is for distributions with high probability for values slightly smaller than the mean, balanced by a low probability for values much larger than the mean.  Lower skewness is the opposite; Fig.~\ref{disrupt}(b) shows an example of low $S$.

In this paper we present computational results of generating random close packing configurations from a wide variety of $P(r)$, allowing us to go beyond prior work which focused on specific distributions such as power law distributions \cite{Shiamomoto,Monti} or bidisperse \cite{Koeze}.  Similar to the 3D results of Desmond and Weeks \cite{Desmond2014}, we find that knowing $\delta$ and $S$ is sufficient to determine the random close packing area fraction $\phi$ to within $\pm 0.002$ in any packing that avoids hexagonal ordering.  Moreover, given that the $\phi(\delta,S)$ observations for a bidisperse size distribution (two distinct circle radii) agree in large part with results for other size distributions, we use analytic results applicable to bidisperse distributions to understand $\phi$ observed in the limits of lowest and highest $S$ (for fixed polydispersity).  Of particular note is that the lowest area fraction we find over all tested distributions is $\phi_0 = 0.840$.  This matches prior results from other groups \cite{Koeze}, although it is higher than the value $0.826$ found for monodisperse circles with a special construction technique \cite{atkinson14}.  We show that the value of $\phi_0$ plays a role in understanding the formula for $\phi(\delta,S)$.

\begin{figure}
    \includegraphics[width=0.22\textwidth]{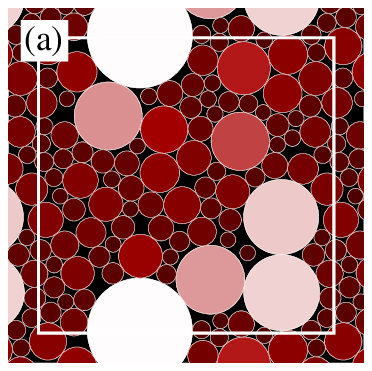}
    \includegraphics[width=0.22\textwidth]{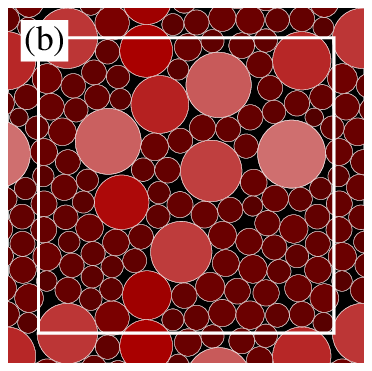}\\
    \includegraphics[width=0.22\textwidth]{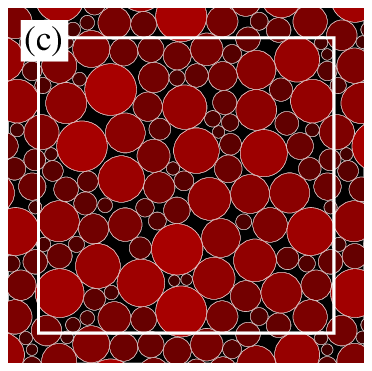}
    \includegraphics[width=0.22\textwidth]{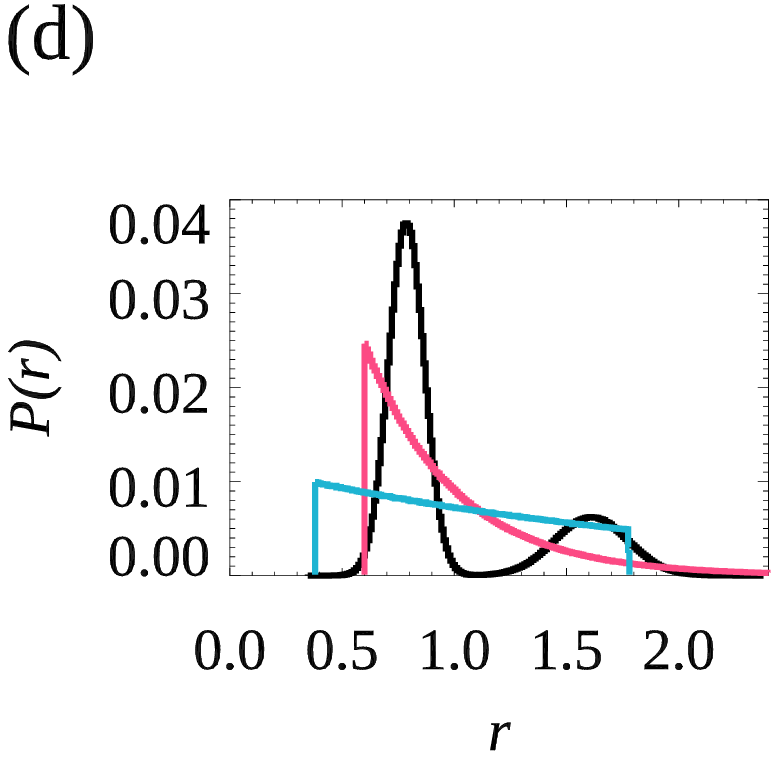}
    \caption{Examples of how the shape of the graph does not matter as much as its statistics. An exponential distribution of sizes [panel (a) and the pink curve in panel (d)] will have the same packing fraction of $\phi\approx 0.86$ as a bidisperse Gaussian distribution [panel (b) and the black curve in d)] if the polydispersity and skewness ($\delta=0.4, \ S=1$) match. An exponential distribution (panel (c) and blue curve in d)] with the same polydispersity but different skewness ($S=0.25$) will pack differently, at $\phi\approx 0.841$. In the packing images the color indicates the relative radius size.  The thin white square indicates the extent of the periodic boundary conditions.  Panel (d) shows the radius distribution functions $P(r)$.}
    \label{shape}
\end{figure}

\section{Protocol}
    \subsection{Simulation}

Our computational methods are a modification of \cite{Desmond2009}, which in turn is based on earlier work by Xu {\it et al.} \cite{xu05} and Clarke and Wiley \cite{clarke87}.  We will describe the algorithm as applied to 2D circle packing.  We start by choosing a circle size distribution $P(r)$ and generate $N$ random radii \cite{numrecipes}.  We choose a starting system size $L$ such that the area fraction is $\phi=\pi \Sigma_i r_i^2 / L^2 = 0.01$.  We then place circles randomly, requiring that no overlaps occur; if placing a given circle would overlap a previously placed circle, we choose a new random position in the $L \times L$ box and try again.  If the size ratio between the smallest and largest circle is more than 50, then we instead using a larger $L$ such that $\phi=0.001$ so that the largest circles can avoid overlapping and keep the initial positions spatially uncorrelated.  We use periodic boundary conditions.

We next shrink the box by a small amount, for example $L_{\rm new} = (1-\alpha) L$ with $\alpha=0.01$, and rescale all circle positions similarly, $x_{\rm new}=(1-\alpha) x$ and $y_{\rm new} = (1-\alpha) y$.  This may cause some circles to overlap.  We assign a potential energy based on a finite range soft interaction potential.  For two circles $i$ and $j$ with a center-to-center separation distance $r_{ij}$ and radii $r_i$ and $r_j$, the interaction potential energy is given by $(r_i + r_j - r_{ij})^2$.  The total potential energy is summed over all overlapping circles.

The circles are then selected in random order.  If a chosen circle overlaps any other circles, we calculate the net force on the circle from its overlapping neighbors and do a one-dimensional minimization of the potential energy by moving that circle in the direction indicated by the net force.  This minimization is halted if a position is found for the circle that does not have any overlaps.  Alternatively, if a chosen circle does not overlap any other circles, then the circle is moved a small step in a random direction as long as this does not cause any overlaps.  The small step size is initially set to be 0.01, and adjusted to smaller values on a per-circle basis for circles that frequently have overlaps when attempting the step.  Allowing these free circles to take this random step results in slightly denser final packings.

After all circles have been moved once, the potential energy is evaluated for the entire system.  If it is zero within numerical tolerance, we again  shrinking the box by the factor $\alpha$ and continue.  If the potential energy is nonzero, then we repeat the attempt to minimize the potential energy up to 50 times.  If the potential energy is still nonzero, we conclude we are at an area fraction where circles must overlap.  In this case, we expand the box by $1/(1-\alpha)$ back to an area fraction where no overlaps are required; decrease $\alpha$ according to $\alpha_{\rm new} = 0.8 \alpha$; and then try shrinking the box by this new factor of $\alpha$.  This lets us approach a state with the smallest box size for which the circles can be placed without overlapping, and thus the highest possible area fraction $\phi$.  As the simulation reaches higher $\phi$ and $\alpha \rightarrow 0$, we occasionally try resetting $\alpha = 0.01$.  The larger value of $\alpha$ results in a state where the forces on overlapping circles are larger and more numerous and sometimes results in circles moving more efficiently, and thus allows the system to reach still higher area fractions without overlapping circles.  We repeat these resetting $\alpha$ trials until the final area fraction changes by less than $0.0005$, at which point we retain the state with the highest area fraction for which there are no overlapping circles, and define this as random close packed.  There are often rattler circles which are not close to touching any other circles; these rattler circles still contribute to $\phi$.

For a given circle size distribution $P(r)$, we repeat this process many times for different numbers of circles $N$. For $N=(100,200,400,800,1600)$ we do $M = (300,140,65,30,15)$ trials.  For a given $N$, we find the mean $\phi(N)$ over the $M$ trials.  We then plot $\phi(N)$ as a function of $N^{-1/d}$ using $d=2$ for circle packing and $d=3$ for sphere packing.  This typically results in data fit well by a straight line, allowing us to extrapolate to $N^{-1/d} \rightarrow 0$ and thus determine $\phi_{rcp}$ for an infinite sized system \cite{Desmond2009}.  For $\delta > 0.8$, the radii distributions often have long tails, so to ensure proper sampling we only use $N \geq 400$ for the extrapolation.

Additionally, the standard deviations from the $M$ trials provide an uncertainty of $\phi(N)$ which lead to an uncertainty of the $N^{-1/d} \rightarrow 0$ intercept, leading to an overall uncertainty of our computed value for $\phi_{rcp}$.  The trial numbers $M$ are chosen to that typically this uncertainty is less than 0.003 and in some cases less than 0.001.

As a first check of our algorithm, we examine three-dimensional results for monodisperse spheres:  our simulation code gives $\phi_{rcp} = 0.6377 \pm 0.0011$.  Desmond and Weeks found $\phi = 0.634$ \cite{Desmond2014}.  Hermes and Dijkstra find $\phi = 0.635 - 0.645$ depending on their protocol \cite{hermes10}.  Our results also agree with a review article which quotes $\phi_{\rm MRJ} \approx 0.64$ \cite{torquato18}, in this case the ``maximally random jammed'' state.  Note that the value of $\phi$ is well-known to depend on computational protocol \cite{torquato18,hermes10,chaudhuri10} so we do not claim our result is universal, but rather provide it to compare with other work.

A highly effective means of estimating $\phi_{\rm 3D}$ without generating actual 3D packings is given by Farr and Groot \cite{Farr}.  However, they note that their method works well in 3D and not in 2D.  This is the main reason we do direct simulation of packings.

\subsection{Choosing circle size distributions}
\label{distributions}

Based on prior work by Desmond and Weeks \cite{Desmond2014}, we conjecture that polydispersity $\delta$ and skewness $S$ are important influences on random close packing of circles.  Therefore we desire to create a variety of circle size distributions $P(r)$ with specific values of $\delta$ and $S$.  To do this, we consider circle size distributions characterized by two parameters which we will label as $a$ and $b$ below.  This then lets us numerically or analytically determine the $(a,b)$ values that give a desired $(\delta,S)$ combination.  In the descriptions below there are also mentions of the mean circle size $\mu$ and a normalization constant $P_0$ which are {\it not} adjustable parameters.  The area fraction of a packing is a nondimensional quantity and thus independent of $\mu$.  Accordingly, we typically define circle size distributions so that $\mu=1$, although occasionally we choose some other convention for numerical convenience, recognizing that $\mu=1$ is not otherwise a requirement.  Likewise, given $a$ and $b$, $P_0$ is chosen so that the integral of $P(r)$ is 1 as required for a probability distribution.

{\it Bidisperse --}
This is a distribution composed of two types of circles with different sizes and specified probabilities of each size.  Desmond and Weeks introduced a mathematical description of bidisperse distributions in terms of two parameters:  the size ratio of the two circles $a = r_-/r_+$, and the probability of one of them $P(r_+) \equiv b$ \cite{Desmond2014}.  In Ref.~\cite{meer24skew} we showed that these two parameters can be replaced by  $\delta$ and $S$ as
\begin{eqnarray}
    r_+ &=& 1+\frac{\delta}{2}\left(S+\sqrt{4+S^2}\right)\label{bS}\\
    r_- &=& 1+\frac{\delta}{2}\left(S-\sqrt{4+S^2}\right)\label{aS}\\
    P(r_+)&=&\frac{\sqrt{4+S^2}-S}{2\sqrt{4+S^2}}\label{Pb}
\end{eqnarray}
These choices ensure $\mu=1$, and thus let us define $P(r)$ for a bidisperse distribution using these analytic formulas. With two circle sizes, $P(r_-) = 1 - P(r_+)$.

{\it Bidisperse Gaussian -- }
As a more realistic realization of physical mixtures of two circles with different mean sizes, we consider a $P(r)$ composed of a sum of two different Gaussians with individual means $r_-$ and $r_+$, with size ratio $a = r_-/ r_+$ and probability $P(r_+) \equiv b$ of finding a circle from the larger of the two species.  The Gaussians have width $0.1r_-$ and $0.1r_+$, that is, $\delta=0.1$ for each individual Gaussian.  To find the values of $a$ and $b$, we scan over a range of these values, considering each $P(r)$ and generating a large number of radii according to the distribution, and computing $(\delta, S)$.  We tune $a$ and $b$ until $(\delta,S)$ is within 0.1\% of the desired values.  For a few cases where we want $0.05 < \delta \leq 0.1$, we use Gaussians with $\delta=0.05$.

{\it Power law -- }
Here $P(r) = P_0 r^a$ for $1 < r < b$.  We consider both $a<0$ and $a>0$ as needed to reach the desired $(\delta,S)$.  Note that $a=0$ is a flat top distribution, where every value of $r$ in the range $1 < r < b$ is equally probable; this achieves $S = 0$.

{\it Exponential -- }
These are distributions $P(r) = P_0 \exp(r / a)$ for $1 < r < b$.

{\it Gaussian -- }
These are distributions $P(r) = P_0 \exp(-(r-1)^2 / a^2)$ for $b < r < c$.  In general we take $c = \infty$ although some distributions with low skewness $S$ require finite values of the third parameter $c$.

{\it Linear -- }
This distribution is defined as $P(r) = Ar + B$ for $r_0 \leq r \leq a r_0$, with a second parameter $b = P(a r_0)/P(r_0)$.  Once parameters $a$ and $b$ are picked, the values of $r_0, A,$ and $B$ are determined by the conditions that the distribution be normalized and that $\langle r \rangle = 1$.  In particular, following \cite{Desmond2014}, we can define $c = 2(b-1)/[(a-1)^2(b+1)],$ $d=2/[(b-1)(a+1)] - c,$ and $e = [c(b^3-1)/3 + d(b^2-1)/2]$, leading to $r_0 = 1/e$, $A=c e^2$, and $B=de$.  Note that a power law distribution with $a=0$ and a linear distribution with $b=1$ are identical flat top distributions with skewness $S=0$.

{\it Parabolic -- }
This distribution is inspired by the observation that two distributions above (bidisperse, bidisperse Gaussian) have two peaks.  Here we define $P(r) = P_0 (r-1)^2$ for $a \leq r \leq b$, thus having one peak at $a<1$ and a second peak at $b>1$.  In this case $\langle r \rangle \neq 1$, although as noted above that is not a strict requirement for $P(r)$.

{\it Tracers in a flat distribution background -- }
We first define a flat top distribution $P(r) = 1/(c_2 - c_1)$ for $c_1 < r < c_2$, with $c_1 = 0.827$ and $c_2=1.173$.  This distribution has mean 1 and $\delta = 0.1$, sufficient to prevent hexagonal ordering as-is.  We then choose radii from this distribution randomly with probability $b$, or choose the radius to be $r=a$ with probability $(1-b)$.

{\it Tracers in $1:1.4$ -- }
This distribution is tridisperse, with circle sizes 1, 1.4, and $a$, and probabilities $(1-b)/2, (1-b)/2,$ and $b$.  In general the idea is that the `$a$' species is a minority, so we prefer $b < 1/3$.  The `majority' is the canonical bidisperse mixture with size ratio $1:1.4$ and equal numbers of circles \cite{perera99,speedy99,ohern03}.  For low skewness typically $a<1$ and for high skewness typically $a>1$; for $S \approx 0$ there are often two solutions $(a,b)$ for a given $(\delta, S)$.  In these cases, for a given $(\delta,S)$ we take the solution with the smaller $b$.

{\it Tracers in $1:1.25$ -- }
This distribution is tridisperse, with circle sizes 1, 1.25, and $a$, and probabilities $(1-b)/2, (1-b)/2,$ and $b$.  The `majority' circles are chosen to match the bidisperse distribution which packs poorly as observed by Koeze {\it et al.} \cite{Koeze}.  This tracer distribution can achieve lower values of $S$ for a given $\delta$ as compared to the tracers in $1:1.4$ distribution.

{\it Quaddisperse --}
This distribution has four particles with size ratios $1:1.25:a:1.25a$ and number ratios $c/2:c/2:b/2:b/2$ with $c = 1 - b$.  This avoids hexagonal ordering for all values of $b$, which sometimes is a problem for the other tracer distributions.

{\it Lognormal -- }
This distribution is defined as
\begin{equation}
    P(r) = \frac{P_0}{r} \exp{[(\ln r)/a + a/2]^2/2}
\end{equation}
where we have only a single parameter $a$.  We choose $a$ to match the desired polydispersity, given by the relation $\delta = \sqrt{\exp(a^2)-1}$.  Once $a$ is known, the skewness is given by $S = [\exp(a^2)+2] \delta$.

{\it Weibull distribution -- }
This distribution is defined as
\begin{equation}
    P(r) = \frac{b}{a} \left( \frac{r}{a} \right)^{b-1} \exp \left[ - \left( \frac{r}{a}\right)^b \right]
\end{equation}
for $r>0$.  $b$ is termed the shape parameter, with $b=1$ producing an exponential distribution.  $a$ is termed the scale parameter.  Both $a$ and $b$ are required to be positive.

\section{Results:  Minimal area fraction packings}
\label{phi0}

We first consider simple circle packings that result in low area fraction random close packed configurations.  One such simple packing is bidisperse, and we start by comparing our results to those of 
Koeze {\it et al.}~\cite{Koeze}.  These authors performed simulations of random close packed configurations using bidisperse size distributions.  Their simulations used a fixed $N=1024$ circles, and they quench their samples at fixed $\phi$ from an initially highly overlapped (completely random) state, looking for the $\phi$ at which half the final states are non-overlapping.  Many people have considered the canonical bidisperse distribution comprised of circles with size ratio $1 : 1.4$ and equal numbers \cite{perera99,speedy99,ohern03}.  For this distribution, Koeze {\it et al.} find $\phi = 0.8394 \pm 0.0002$.  We interpolate our results to find $\phi(N=1024) = 0.8394 \pm 0.0003$.  This striking agreement suggests that despite the different algorithms, our methods find similar states.  Our extrapolated value for an infinite size system is $\phi(N \rightarrow \infty) = 0.8419 \pm 0.0003$.  Koeze {\it et al.} also identify a different system with the lowest value of $\phi$; this system has equal numbers of small and large circles and size ratio $1 : 1.25$.  They do not state the value of $\phi$; for this state we find $\phi = 0.8402 \pm 0.0003$, which is indeed slightly smaller than our infinite system size result for the $1 : 1.4$ size ratio.  It is this size distribution that we use as the ``background'' mixture of the third tracer distribution described in the previous subsection.  Finally, Koeze {\it et al.} identify one other local minimum of $\phi$ within the parameter space they scan, with 20\% small circles and 80\% large, size ratio $1 : 2.5$.  We find $\phi = 0.8471 \pm 0.0004$ for this system.

\begin{figure}
    \centering
    \includegraphics[width=0.45\textwidth]{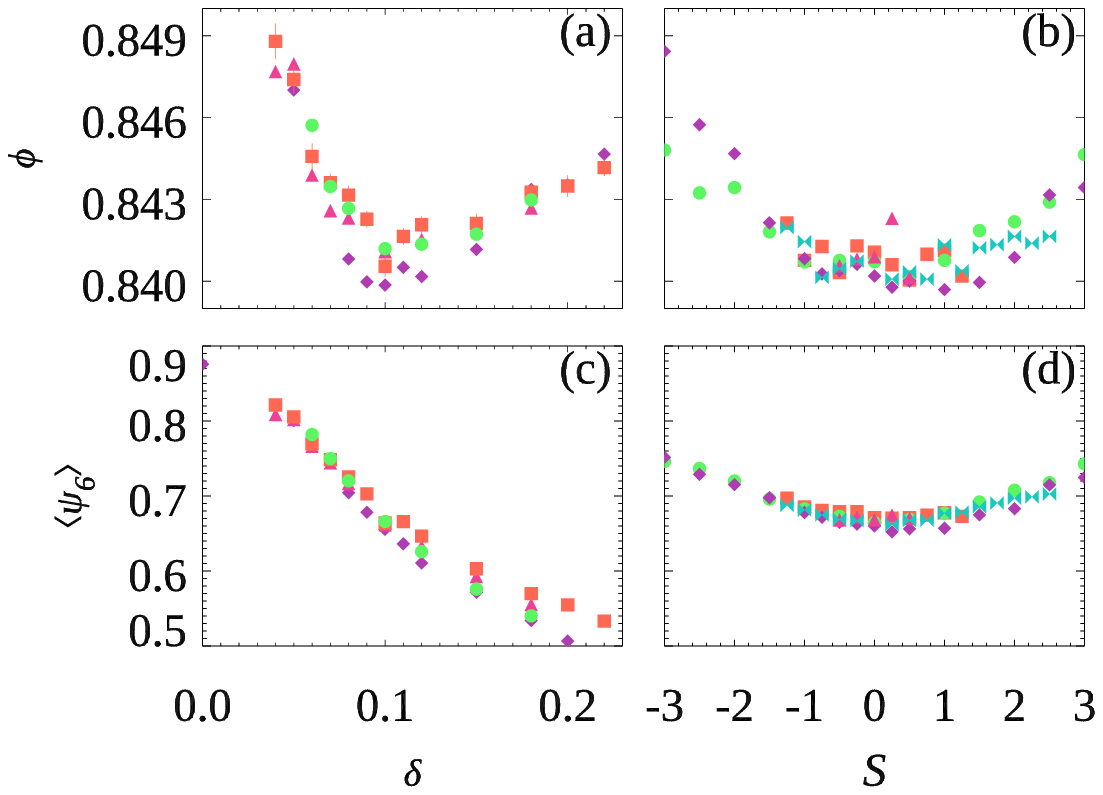}
        \caption{(a) Random close packing area fraction $\phi$ as a function of $\delta$ for symmetric radii distributions $P(r)$.  The distributions are bidisperse (diamonds), Gaussian (squares), bidisperse Gaussian (circles), and linear (triangles).  (b) $\phi$ as a function of $S$ for distributions with $\delta = 0.1$.  The symbols are as in (a), along with power law distributions (bow ties).  (c) Sample averaged hexagonal order parameter $\langle \psi_6 \rangle$ as a function of $\delta$ for the data shown in (a).  (d) $\langle \psi_6 \rangle$ as a function of $S$ for the data shown in (b).}
    \label{phi10}
\end{figure}

Across the array of simulations, the results for the bidisperse $1 : 1.25$ mixture is nearly the lowest value of $\phi$ we found.  The study of Koeze {\it et al.} just considered bidisperse distributions, but of course there are other circle size distributions that might achieve a low area fraction.  In particular, one wishes to avoid hexagonal ordering (which would increase $\phi$) so some polydispersity is needed.  Accordingly, we examine $\phi$ for a variety of symmetric radii distributions $P(r)$ with modest polydispersity $\delta$.  The results of $\phi(\delta)$ are shown in Fig.~\ref{phi10}(a) for four different distributions.  There is a clear minimum at $(\delta \approx 0.10, \phi \approx 0.840)$.  For the bidisperse distribution, $\delta = 0.10$ requires equal numbers of small and large circles with size ratio $1 : 1.222$, quite close to the $1 : 1.25$ minimal $\phi$ mixture identified by Koeze {\it et al.} \cite{Koeze}.

Figure \ref{phi10}(a) shows results for symmetric distributions ($S = 0$) and we wish to understand how slight skewness influences $\phi$.  We fix polydispersity $\delta = 0.1$ and vary $S$, giving the results shown in Fig.~\ref{phi10}(b).  $\phi$ does not strongly depend on $S$, but what dependence there is suggests that $S=0$ minimizes $\phi$.  Overall, we conjecture that any circle size distribution which is symmetric ($S=0$) and achieves polydispersity $\delta = 0.1$ will lead to a random close packing configuration with $\phi_0 \approx 0.840$.  This includes the bidisperse distribution with size ratio $1 : 1.222$, a Gaussian with $\delta = 0.1$, and a linear distribution with $a = 1.419, b=1$ (flat top distribution).

At low polydispersity, hexagonal ordering can increase the packing fraction. To measure this we compute the hexagonal order parameter $\psi_6$.  The starting point for this order parameter is to define nearest neighbor circles through the Delaunay triangulation.  This triangulation method connects the centers of circles in a unique tiling of triangles.  In particular, each triangle so formed is constructed so that no other circle centers are within the circumcircle of the triangle.  This triangulation connects each circle to its nearest neighbors.  Relative to an arbitrary $x$-axis, the $N_k$ nearest neighbors $k$ of circle $j$ are oriented at angles $\theta_{jk}$.  $\psi_6$ is then defined as
\begin{equation}
    \psi_6 = \frac{1}{N_k} \Sigma_k \exp(6 i \theta_{jk}) 
\end{equation}
where $i = \sqrt{-1}$.  For our purposes we consider just the magnitude $|\psi_6|$ which is between 0 and 1.  $|\psi_6 |= 1$ indicates perfect hexagonal ordering, where all the $\theta_{jk}$ are separated by multiples of $\pi/3$ radians.  From this point we drop the absolute value signs, referring to this  real number as $\psi_6$.  Figure \ref{disrupt} shows two examples of packed circles colored by their $\psi_6$ value with white regions in the left image being nearly perfectly hexagonal, and darker green regions in the right image being quite different from hexagonal.

For the low polydispersity data of Fig.~\ref{phi10}(a,b), the sample averaged $\psi_6$ are plotted in Fig.~\ref{phi10}(c,d).  The larger values of $\phi$ at low $\delta$ seen in Fig.~\ref{phi10}(a) are due to hexagonal ordering.  The minimum in $\phi$ at $\delta = 0.1$ appears to be a tradeoff between having enough polydispersity to decrease hexagonal ordering, without too much polydispersity so that smaller circles efficiently fill in voids between larger ones.  The minimum in $\phi$ seen at $S\approx 0$ in Fig.~\ref{phi10}(b) coincides with a minimum in $\langle \psi_6 \rangle$ seen in Fig.~\ref{phi10}(d), showing that the symmetric distributions minimize hexagonal ordering, helping them achieve a minimal $\phi$.

At the limit of monodisperse circles, we find $\psi_6 = 0.880 \pm 0.010$.  The value is less than one due to random defects and grain boundaries.  The area fraction we find for monodisperse circles is $\phi = 0.862 \pm 0.002$, less than the ideal packing $\phi_{\rm ideal} = \pi / (2 \sqrt{3}) \approx 0.907$ again due to the defects and grain boundaries.  An example of such a packing is shown in Fig.~\ref{disrupt}(a).

While we cannot rule out the existence of some circle size distribution that achieves a still lower $\phi$, it seems plausible given our results that $\phi_0 = 0.840$ is a lower bound on random close packing in 2D.  An easy way to achieve this packing is the bidisperse mixture, equal numbers of small and large disks, with size ratio $1 : 1.222$ (with the latter number equal to 11/9 to achieve $\delta = 1/10$, as per Eqs.~\ref{bS}-\ref{Pb}).  However, note that mechanically stable disordered packings have been found with lower area fractions using special techniques \cite{atkinson14,dennis22}.

\section{Results:  Higher polydispersity} \label{theory}

\begin{figure}
    \centering
    \includegraphics[width=0.45\textwidth]{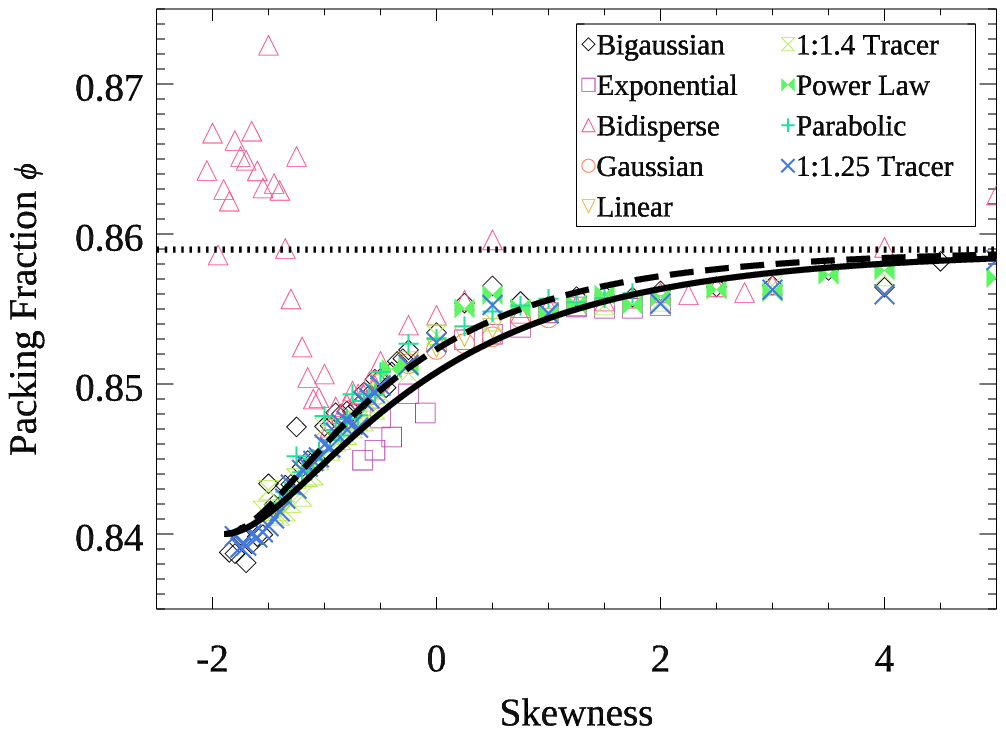}
\caption{The packing $\phi$ as a function of skewness for $\delta$=0.4, with the analytical fit line (solid) and least squares fit line (dashed) discussed in section \ref{fitting}.  Both fit curves asymptote to the dotted line which indicates $\phi_1$. The bidisperse data (upward triangles) deviates at low skewness due to hexagonal ordering. }
    \label{crystalrcp}
\end{figure}

As described above, we choose a variety of values for the polydispersity and skewness $(\delta, S)$, choose appropriate radii distributions that achieve those values, and simulate to find the area fractions $\phi$ for each case.  Figure \ref{crystalrcp} shows the results for a representative polydispersity $\delta = 0.4$.  Ten different types of radii distributions are shown, and the majority of the data collapses fairly well onto a master curve.

\subsection{Small skewness and hexagonal ordering}

\begin{figure}
    \centering
    \includegraphics[width=0.45\textwidth]{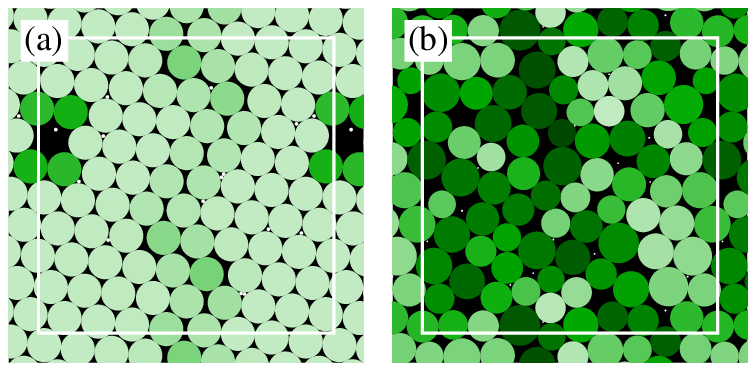}
    \caption{
    Two packings with $\delta = 0.4$ and $S=-1.75$.  (a) This bidisperse sample packs at  $\phi\approx 0.87$, with a high degree of order. (b) A bidisperse Gaussian packing with $\phi\approx\phi_0$. For both images particles are colored according to their $\psi_6$ value, with lighter green corresponding to higher $\psi_6$.  The small white particles are ignored in the calculation of $\psi_6$.}
    \label{picdel40}
\end{figure}

The data in Fig.~\ref{crystalrcp} can be traced to as large a positive skew as desired, but it is impossible to find distributions below $-2.1$ for this polydispersity.  This has to do with the requirements for low skewness distributions, which requires a large number of large circles, and only a small number of small circles -- and the size of the small circles needs to be especially small, see for example the two packings shown in Fig.~\ref{picdel40}.
In fact, for a bidisperse sample the smallest size of the small circles is $r_{-} \rightarrow 0$:  circles cannot have negative radius, but their radius can be arbitrarily small.  In this case, Eq.~\ref{aS} can be solved for $r_{-}=0$, leading to the minimum possible skewness
\begin{equation}
    S_0 = \frac{\delta^2-1}{\delta}. \label{S0}
\end{equation}
While this is derived from the bidisperse distribution, this is in fact the mathematical lower bound for skewness for any distribution $P(r)$ subject to the constraint $r \ge 0$, and this bound is only achieved using a bidisperse distribution \cite{smolalski20,meer24skew}.  For $\delta = 0.4$, we have $S_0 = -2.1$, and this is indeed the smallest skewness value shown for the bidisperse data in Fig.~\ref{crystalrcp}.

For the bidisperse distribution (triangles) as $S \rightarrow S_0$, the limit is the packing of a monodisperse sample with some hexagonal order, which as noted previously is $\phi \approx 0.862$ for our simulation protocol.  The scatter around this value seen in Fig.~\ref{crystalrcp} for the triangle symbols reflects the larger uncertainty for these nearly hexagonal packings; variations in the density of defects and grain boundaries results in large fluctuations of $\phi$.  The hexagonal packing results in an increased $\phi$ for the bidisperse data for $S \lesssim -1$.  For sufficiently small $S$, the smaller sized circles are small enough to fit into the voids between the large circles, allowing the more numerous large circles to organize into hexagonal arrays.

\subsection{Small skewness and amorphous packing}

Examining the distributions other than the pure bidisperse, the data of $\phi(S)$ collapse remarkably well in Fig.~\ref{crystalrcp} across the variety of radius distributions.  The reason for this collapse is unknown, although it is similar to what was observed by Desmond and Weeks for 3D packings \cite{Desmond2014}.  That being said, given that the data collapse, we can understand some features of the master curve by considering specific distributions.
The violet diamonds in Fig.~\ref{crystalrcp} correspond to bidisperse Gaussian distributions, chosen to match $\delta=0.4$ (for this figure) and the desired $S$.  As with the pure bidisperse, the bidisperse Gaussian reaches the most negative value of skewness when the smaller circle mean size goes to zero.  

Only bidisperse distributions allow $S$ to reach $S_0$. For the bidisperse Gaussian case the large circle species is not a single size but rather a Gaussian with polydispersity $\delta_i=0.1$.  This modifies the minimum possible skewness to be bounded by 
\begin{equation}
S_1=\frac{\delta^{4}\left(1+3\delta_{i}^{2}\right)-\delta^{2}\left(1+3\delta_{i}^{4}\right)+\delta_{i}^{2}\left(1-\delta_{i}^{2}\right)}{\delta^{3}\left(1+\delta_{i}^{2}\right)^{2}} \label{s0'}
\end{equation}
with the derivation of this equation given in Appendix A.  In fact, this result for $S_1$ is not specific for the large circle species being described by a Gaussian, but rather, it is correct for any distribution function for the large circle species having polydispersity $\delta_i$ and initial skewness $S_i = 0$, which you combine with a delta function centered at $r=0$.  For the bidisperse-Gaussian data presented in Fig.~\ref{crystalrcp} (violet diamonds), we have $\delta = 0.4$ and $\delta_i=0.1$, leading to $S_1 \approx -1.896$ which is slightly larger than $S_0 = -2.1$.  This then explains the lowest $S$ data plotted for the bidisperse Gaussian results; one cannot achieve a lower value of $S$ with this distribution type.

Thus we understand the lower left corner of the master curve of Fig.~\ref{crystalrcp}:  as $S \rightarrow S_1$, the system is composed of a large circle species represented by a Gaussian distribution with $\delta_i = 0.1$, and a small circle species that becomes negligible in size.  The packing is similar to the sketch in Fig.~\ref{phi12}(b) where the small particles do not affect the packing structure, but rather exist as rattlers in the voids between the large particles.  The data in this limit have $\phi \rightarrow \phi_0$, our minimum area fraction which is exactly that found by distributions with $(\delta_i = 0.1, S = 0)$ as discussed in Sec.~\ref{phi0}.  Thus the lower corner of the master curve {\it must} be at the point $(S_1,\phi_0)$ as no lower $S$ is possible without choosing $\delta_i$ in such a way that would increase the packing from $\phi_0$.  The general overlapping of the data at higher $S$ in Fig.~\ref{crystalrcp} is intriguing, as many of the distributions we simulate cannot reach $S_1$ yet seem to fall onto the same curve which is heading toward $(S_1, \phi_0)$.

\subsection{High skewness}
\label{highskew}

The data of Fig.~\ref{crystalrcp} appear to be heading to an asymptote for large $S$.  This asymptotic limit can be understood by considering pure bidisperse distributions.  We expressed the sizes $(r_+,r_-)$ of bidisperse circles in terms of polydispersity and skewness in Eqs.~\ref{bS} and \ref{aS}.  Equation \ref{Pb} quantifies how often each particle size appears [$P(r_+)=1-P(r_-)]$, allowing us to calculate the relative proportion $N$, the total number of small circles in a system per one large one:
\begin{eqnarray}
    N&=&\frac{P(r_-)}{P(r_+)}=\frac{\sqrt{4+S^2}+S}{\sqrt{4+S^2}-S}\label{NS}
\end{eqnarray}
As skewness grows large, to leading order the results of Eqs.~\ref{bS}, \ref{aS}, and \ref{NS} become
\begin{eqnarray}
    r_+ &\approx& \delta S\\
    r_- &\approx& 1 \\
    N &\approx& S^2.
\end{eqnarray}
Thus, for large $S$ we arrive at a bidisperse system with many small circles for each large circle.

\begin{figure}
    \centering
    \includegraphics[width=0.4\textwidth]{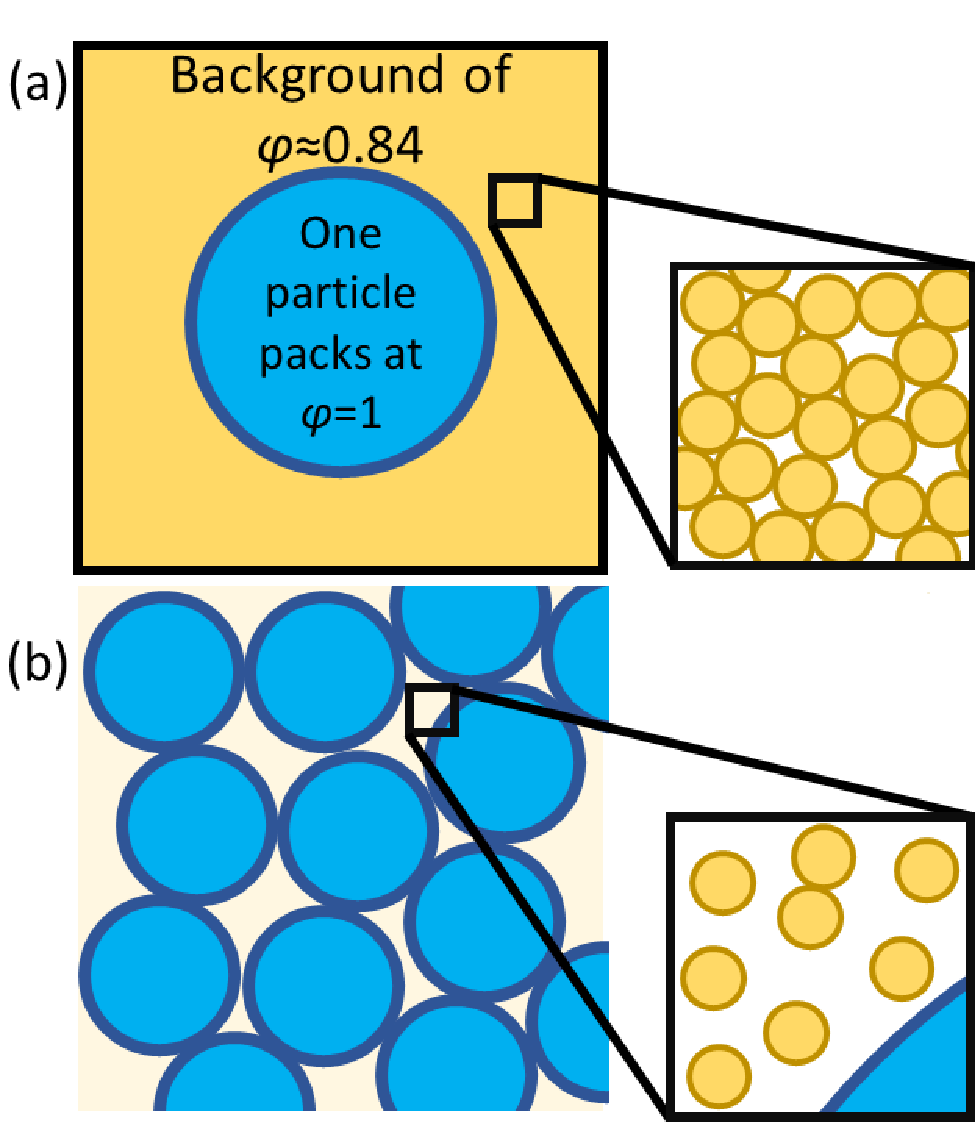}
    \caption{Cartoons of the theoretical limits of $f_1$ (a) and $f_2$ (b). In $f_1$, the background field of circles pack randomly, and are disrupted by a large circle packing at $\phi=1$. In $f_2$, the large circles pack randomly, and the small circles do not disrupt the packing or fill in the voids.}
    \label{phi12}
\end{figure}

To compute the area fraction in this situation, note that for $\delta \leq 1$ the small circles are so numerous that they dominate the area of the system; see Fig.~\ref{phi12}(a).  Nonetheless, the large circles are still frequent and large enough to contribute to the overall area fraction.  A large circle is essentially its own region with $\phi=1$.  Assuming a perfectly bidisperse system, the ``sea'' of small circles is monodisperse and would form hexagonal patches.  To consider only random packings, we consider instead a situation where the small circles have a mean radius of $r_-$ and a polydispersity $\delta_i = 0.1$, as is the case for the bidisperse Gaussian and our ``tracers in $1:1.25$'' distribution.  Given that $r_+ / r_- \gg 1$, it is reasonable to approximate the small circles as still following the bidisperse formulas while packing at area fraction $\phi_0$.  Therefore, Fig.~\ref{phi12}(a) suggests a formula for the packing fraction $\phi_1$ of this system by computing the total area of circles divided by the area occupied:
\begin{eqnarray}
    f_1(\delta, S) &=&\frac{Nr_-^2+r_+^2}{(Nr_-^2/\phi_0)+r_+^2}\nonumber\\
    \lim_{S\to\infty} f_1(S) &=&\phi_1(\delta)=\frac{\phi_0+\delta^2\phi_0}{1+\delta^2\phi_0}.\label{phi1}
\end{eqnarray}
Indeed this matches the asymptotic behavior of the large $S$ data of Fig.~\ref{crystalrcp}, with $\delta = 0.4$ and $\phi_0 = 0.840$ leading to $\phi_1 \approx 0.861$.  The slight increase in the squares at $S \geq 4$ seen in Fig.~\ref{crystalrcp} is because for the bidisperse data, the background ``sea'' is starting to form hexagonal order, emphasizing the utility of the other distributions for suppressing hexagonal ordering at high skewness.

\begin{figure}
    \centering
    \includegraphics[width=0.49\textwidth]{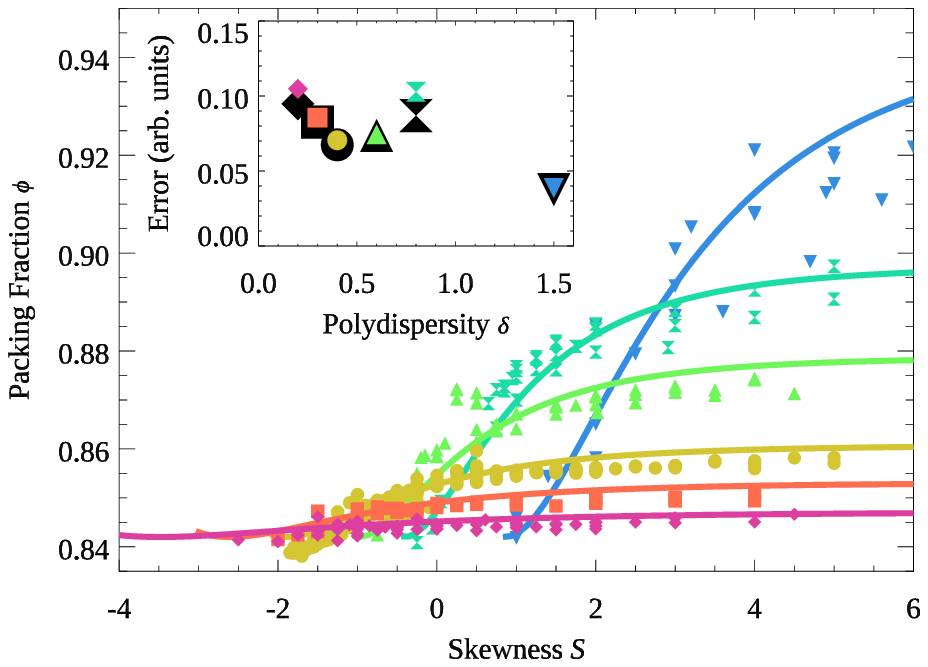}
    \caption{The data for the packing fraction for many different circle size distributions with polydispersity from bottom to top given by $\delta= 0.2$ (pink diamonds), 0.3 (orange squares), 0.4 (yellow circles), 0.6 (green upward triangles), 0.8 (cyan hourglasses), and 1.5 (blue downward triangles). The lines are best fits to Eq.~\ref{lognorm}.
    (Inset) shows the fitting errors defined in Eq.~\ref{erroreqn}.  The large black symbols are calculated from the least squares fitting for $p$, and the smaller colored symbols at slightly higher error values are from the analytic expression for $p$ using Eq.~\ref{panalytic}.}
    \label{alldelts}
\end{figure}

All of this has been discussed in the context of Fig.~\ref{crystalrcp} which is for the specific polydispersity $\delta = 0.4$.  The trends are similar for other polydispersities as shown in Fig.~\ref{alldelts}:  a lower left corner at $[S_1(\delta),\phi_0]$ and an asymptote at $f_1(\delta)$ for large $S$.  The exception to these results is the $\delta = 0.1$ data shown in Fig.~\ref{phi10}(b), which as previously discussed reaches the minimum $\phi_0$ at $S=0$ and otherwise has larger $\phi$ due to hexagonal ordering.

We note that there is a packing exactly intermediate between the two cases of Fig.~\ref{phi12}:  where the large particles are random close packed at $\phi_0$, and likewise the small particles are also random close packed at $\phi_0$ in the voids between the large particles.  This ideal intermediate case packs at $\phi_0+(\phi_0)(1-\phi_0)=2\phi_0-\phi_0^2 = 0.9744$. Setting $\phi_1 = 2\phi_0-\phi_0^2$ in Eq.~\ref{phi1} and solving for $\delta$ gives $\delta = \frac{1}{\sqrt{1-\phi_0}}$.  If our minimum packing is $\phi_0\approx 0.840\pm 0.001$, this gives a polydispersity of maximum packing $\delta'\approx 2.500 \pm 0.008$.  Of course, achieving this packing requires the small particles to allocate in correct proportions to the voids, which is improbable. 

The main point is that for $\delta \gtrsim 2.5$, we expect the packing to cross over to the situation shown in Fig.~\ref{phi12}(b) with large particles packing at $\phi_0$ and small particles existing as rattlers in the voids.  The packing fraction in this situation is given by
\begin{eqnarray}
    f_2(\delta,S) &=&\frac{Nr_-^2+r_+^2}{(r_+^2/\phi_0)}\label{f2eqn}\\
    \lim_{S\to\infty} f_2(\delta,S) &=&\phi_2(\delta)=\left(1 + \frac{1}{\delta^2}\right) \phi_0.\label{phi2}
\end{eqnarray}
In the $S\to\infty, \delta\to\infty$ limit, the small particles become negligible and the packing fraction becomes  $\phi_0$,  similar to the behavior noted at $S\to S_1$ for any $\delta$.

Of course, Eq.~\ref{phi2} gives the asymptotic behavior.  One can imagine packings constructed with specific size distributions designed to pack more densely than bidisperse.  Consider a tridisperse system where $r_1 \gg r_2 \gg r_3$. The smallest particles can fill in the voids for a medium particles, which in turn fill in the voids for the largest particles. Therefore, the pack can exceed $2\phi_0-\phi_0^2$, the ``maximum'' deriving from bidisperse particles, but this will have specific finite values for $\delta$ and $S$ depending on the details.  Simulating highly polydisperse and skewed particle distributions, and the massive amount of particles that requires to avoid finite-size effects, is prohibitively expensive in computation power, so we leave this as an open question for future work.  For that matter, we have limited our simulations to $\delta \leq 1.5$ due to similar considerations, so our data are not at high enough $\delta$ to see a large $S$ asymptote to $\phi_2$.

\subsection{Empirical Fit} \label{fitting}

\par
\begin{figure}
    \centering
    \includegraphics[width=0.49\textwidth]{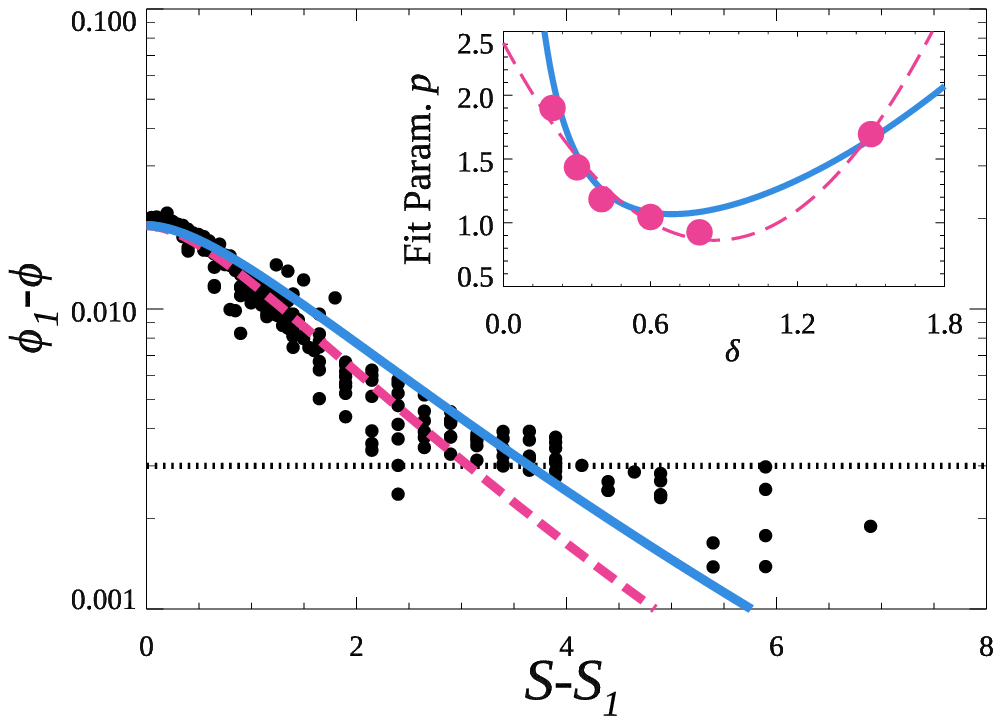}
    \caption{The data of Fig.~\ref{crystalrcp} replotted as $\phi_1 - \phi$ as a function of $S-S_1$ to highlight the approach to the asymptotic limit at large $S$.  The lines are the analytical fit for $p$ (blue solid line, Eq.~\ref{panalytic}); the least squares fit (pink dashed line); and the measurement error for $\phi$ (black dotted). (Inset) $p$ fit parameter (circles) for many $\delta$ with the analytical curve (solid blue, Eq.~\ref{panalytic}) and least squares parabolic fit to the data (dashed pink, Eq.~\ref{peqn}).}
    \label{semilogplot}
\end{figure}

 As discussed, the master curve of $\phi$ as a function of $S$ for fixed $\delta$ has a characteristic shape with known results as $S \rightarrow S_1$ and $S \rightarrow \infty$.  However, the data are not clear as to what functional form bridges these two limits.  To investigate this, the data of Fig.~\ref{crystalrcp} are replotted in Fig.~\ref{semilogplot}, which highlights the behavior as $S \rightarrow \infty$ and thus $\phi \rightarrow \phi_1$.  The horizontal dashed line indicates the uncertainty of the $\phi$ data; the points below this line are such that $\phi_1 - \phi$ is indistinguishable from zero.

We note that Eq.~\ref{phi2} can be taken in the limit $S \rightarrow S_1$ where it can be seen that as $S$ increases from a low value, the contribution of the small particles to the area fraction rises quadratically (until the small particles are large enough to perturb the packing of the large particles).  This suggests a function form that grows quadratically at small $S$ and slowly asymptotes at large $S$, with a nearly-linear regime in the middle section. The log-normal function satisfies these suggestions, so we fit our data to 
\begin{equation}
    \phi_{\rm LN} = \phi_1 - (\phi_1-\phi_0) \exp\left[ -(\ln x)^2\right]
    \label{lognorm}
\end{equation}
with $x \equiv (S - S_1 + p)/p$, and $p$ is the sole fitting parameter.  This works well, as seen by the dashed lines in Figs.~\ref{crystalrcp} and \ref{semilogplot}.  The dependence of $p$ on $\delta$ is given by
\begin{equation}
    p \approx 2.1(\delta - 0.86)^2 + 0.86
    \label{peqn}
\end{equation}
An alternative way to derive $p$ is to require the concavity at $S_1$ in Eq.~\ref{peqn} be equal to the concavity of the continuous packings at $S_1$. This can be accurately approximated by using Eq.~\ref{f2eqn} to determine the concavity of $f_2(S_0)$, which is more easily computed. Setting the concavities equal leads to the following equation for $p$:
\begin{equation}
    p = \frac{\left(\delta^{2}+1\right)^{2}}{\delta}\sqrt{\frac{1-\phi_{0}}{1+\delta^{2}\phi_{0}}}
    \label{panalytic}
\end{equation}
Equation \ref{panalytic} creates a natural, parameterless fit shown by the solid lines in Figs.~\ref{crystalrcp} and   \ref{semilogplot}.  This predicted fit is quite close to the least squares fit (dashed lines).  The inset to Fig.~\ref{semilogplot} shows the least-squares fitting values of $p$ (symbols), the approximation to $p$ given by Eq.~\ref{peqn} as the dashed line, and the analytic result given by Eq.~\ref{panalytic} as the solid line.  It is clear that Eq.~\ref{panalytic} describes the results for $p(\delta)$ nearly as well as the empirical quadratic fit.

We quantify the fitting quality by computing a least squares error and normalizing by the total theoretical range of our pack, $\phi_1-\phi_0$.  Thus the error is given as
\begin{equation}
    \sigma=\frac{\sqrt{\frac{1}{N}\sum(\phi_{\textrm{sim}}-\phi_{\textrm{fit}})^2}}{\phi_1-\phi_0}
    \label{erroreqn}
\end{equation}
The errors from this equation are plotted in the inset of Fig.~\ref{alldelts} for the situation where $p$ is set by Eq.~\ref{panalytic} using the smaller colored symbols, and for the situation where $p$ is determined by minimizing $\sigma$ using the larger black symbols.  By construction the black symbols are the best case for $\sigma$, but nonetheless the errors using the analytic result for $p(\delta)$ are almost as good.  Accordingly, we suggest Eq.~\ref{lognorm} with Eq.~\ref{panalytic} is the best method to estimate $\phi$ based on polydispersity $\delta$ and skewness $S$ for a new particle size distribution, using $\phi_0=0.840$.  We note that we do not have certain knowledge that the log-normal function (Eq.~\ref{lognorm}) is the best way to describe the data, but as noted above, it is the simplest function that satisfies the obvious characteristics of the data.

\section{Implications for 3D packings}

\begin{figure}
    \centering
    \includegraphics[width=0.45\textwidth]{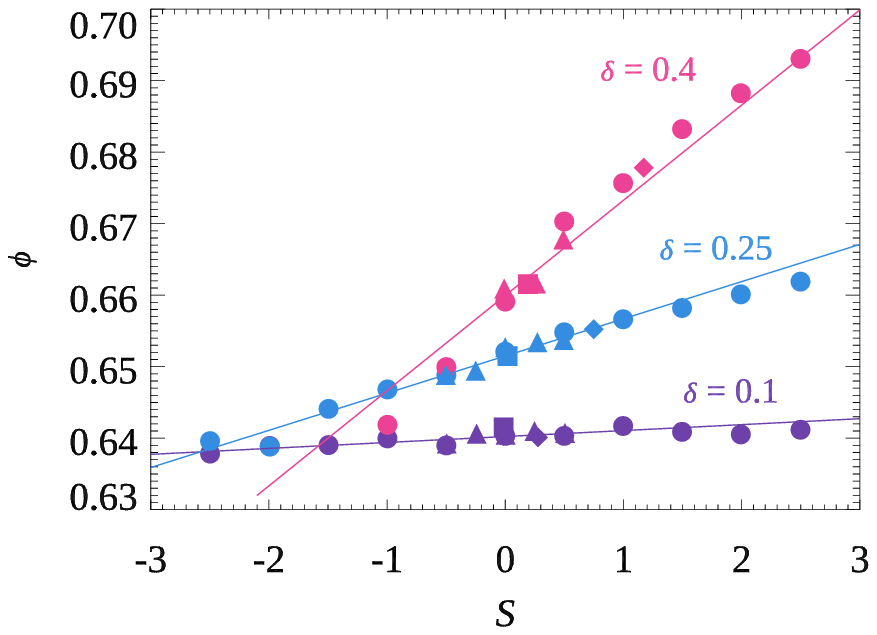}
            \caption{Replotted data from Desmond and Weeks \cite{Desmond2014} showing 3D volume fraction $\phi$ as a function of skewness $S$ for values of polydispersity $\delta$ as indicated.  The circles indicate bidisperse distributions, the triangles indicate linear distributions, the squares are Gaussian distributions, and the diamonds are log-normal distributions.  The lines are a two parameter fit of the form $\phi(\delta,S) = \phi_{\rm rcp}^* + a \delta^2 (S-S_0)$ with $\phi_{\rm rcp}^*=0.632$, $a=0.0832$, and $S_0$ given by Eq.~\ref{S0}.}
    \label{kendata}
\end{figure}

Our understanding of the 2D results suggest a new interpretation for the 3D data of Desmond and Weeks \cite{Desmond2014}, who found that $\phi_{\rm 3D}$ was a function of $\delta$ and $S$.  Their data are replotted in Fig.~\ref{kendata}.  In Ref.~\cite{Desmond2014} they provided a fitting function for their data with three fit parameters:
\begin{equation}
    \phi_{\rm 3D}(\delta,S) = \phi_{\rm rcp}^* + c_1 \delta + c_2 S \delta^2
    \label{kenformula}
\end{equation}
where $\phi_{\rm rcp}^* = 0.634$ is the packing fraction for monodisperse circles, and $c_1 = 0.0658$ and $c_2 = 0.0857$ are empirical constants.  Our new insight is that for any $\delta$, one can consider a bidisperse distribution with the smallest circle size reaching a limit of zero, and thus $S \rightarrow S_0$ with the minimum skewness $S_0$ given by Eq.~\ref{S0}.  In this limit $\phi_{\rm 3D} \rightarrow \phi_{\rm rcp}^*$.  We refit the data plotted in Fig.~\ref{kendata} using the constraint of $\phi_{\rm 3D}(\delta,S_0) = \phi_{\rm rcp}^*$ as a fixed value, and find a new fitting function with only two fitting parameters:
\begin{equation}
    \phi_{\rm 3D}(\delta,S) = \phi_{\rm rcp}^* + c (S-S_0) \delta^2    
    \label{newkenformula}
\end{equation}
with the new value $\phi_{\rm rcp}^* = 0.632$ and parameter $c = 0.0832$.  Plugging in Eq.~\ref{S0} into Eq.~\ref{newkenformula} shows that Eqs.~\ref{kenformula} and \ref{newkenformula} are nearly the same for the values of $\delta \leq 0.4$ considered in Ref.~\cite{Desmond2014}.  This new fit gives the solid lines shown in Fig.~\ref{kendata}.

However, our analysis of the bidisperse distribution reveals that the linear dependence of $\phi_{\rm 3D}$ on $S$ cannot be strictly true.  For the smallest values of $S$ approaching $S_0$, assuming a bidisperse distribution, the small circles are much smaller than the voids between the large circles, the 3D equivalent of Fig.~\ref{phi12}(b).  In this situation, the small spheres contribute their volume to the volume fraction but do not otherwise perturb the structure.   It is straightforward to calculate their contribution.  Consider a cube with edge length $L$ packed randomly with $N_+$ large spheres at $\phi_{\rm rcp}^*$ and with $N_-$ small spheres in the voids.  By construction, $L^3 = N_+ (\frac{4}{3}\pi r_+^3) \phi_{\rm rcp}^{-1}$.  Assuming the small species easily sit within the voids, the volume fraction contribution of the small species is $\phi_- = N_- (\frac{4}{3}\pi r_-^3) L^{-3} = (N_- / N_+)(r_-^3/r_+^3) \phi_{\rm rcp}$.  The ratio $N_-/N_+$ is equal to the ratio of the probabilities of the two circle sizes $P_-/P_+$.  From Eqs.~\ref{bS}, \ref{aS}, and \ref{Pb}, it can be worked out that for $S \approx S_0$, $P_-/P_+ \approx \delta^2$, and $r_-$ grows from zero linearly with $(S-S_0)$, to lowest order in $(S-S_0)$.  Thus $\phi_-$ grows approximately as $(S-S_0)^3$.  This demonstrates that the linear trend $\phi \sim S$ is only roughly correct over the larger range of $S$ shown in Fig.~\ref{kendata} and qualitatively incorrect as $S \rightarrow S_0$.

\section{Conclusion}

We have computationally generated a large number of close packed states from a wide variety of radius distribution functions.  In some cases where there are many particles of similar size, hexagonal ordering occurs and the area fraction $\phi$ can be fairly large.  In cases where hexagonal ordering is avoided, we find the area fraction is well predicted by the polydispersity $\delta$ and skewness $S$ of the radius distribution function.  In particular, the analytic description of the master curve is given by Eqs.~\ref{lognorm} and \ref{panalytic} using our observed minimum possible random close packing area fraction $\phi_0=0.840$.  These results allow one to predict the random close packing area fraction for any radius distribution function, so long as there is not significant hexagonal ordering.  Apart from the polydispersity and skewness, the prediction is independent of the underlying shape of the radius distribution function.

A radius distribution has an infinite number of moments, so it is intriguing that knowing just two of them ($\delta$ and $S$)  collapse our $\phi$ data independent of distribution type.  We do not know why this collapse works so well.  Nonetheless, the observation of the agreement between various radius distributions with matched $\delta$ and $S$ allows us to understand the shape of the master curve by considering the simplest distributions which obey it.  In particular, the bidisperse radius distributions and bidisperse Gaussian distributions are mathematically useful.  However, we stress that our results apply to a broader range of radius size distributions including for example power-law distributions for which radii cover a wide range of values without any bimodal character.

The bidisperse Gaussian distribution mixes together two distinct species, each of which is described by a Gaussian.  One can imagine generalizations of the idea of mixing two species.  The minimal packing fraction $\phi_0$ could be replaced by the observed packing fraction of a single species.  For example, one could use the results of Fig.~\ref{phi10} which show the area fraction for single particle species of low to moderate polydispersity.  Alternatively, one might consider random loose packing where $\phi_0$ is replaced by a smaller number.  Having a new value of $\phi_0$ would then allow one to follow the reasoning outlined in Sec.~\ref{highskew} and use Eqs.~\ref{lognorm} and \ref{panalytic} to predict random close packing for mixtures of two species.  In this broader sense, these two equations represent a prediction for random close packed samples with a sole fitting parameter, $\phi_0$, which would allow one to generalize our results to different computational algorithms or potentially differing experimental conditions.

\begin{acknowledgments}
We thank Z.~Germain and S.~Hilgenfeldt for helpful discussions.  This material is based upon work supported by the National Science Foundation under Grant No. CBET-2306371.  JGFM acknowledges Coordination for the Improvement of Higher Education Personnel (CAPES) - Financing Code 001 for his scholarship

\end{acknowledgments}

\section*{Appendix A: Derivation of $S_1$}
We wish to derive information about an initial distribution $P_i(x)$ with mean $\mu=1$ added with some weight $\alpha$ to a Dirac delta function $\Delta(x)$ at $x=0$ at weight $1-\alpha$, to some new combined $P_c(x)$. This delta function implies the behavior mimics a limit as skewness approaches its minimum. The initial function has polydispersity and skewness $\delta_i,S_i$. As we modify $\alpha$, we want to find the lowest possible skewness of the entire system, for a given total polydispersity. Our combined function is therefore given as:

\begin{equation}
P_c(x)=\alpha P_i(x)+(1-\alpha)\Delta(x)
\end{equation}
This equation implies an average $\mu=\alpha$. Let us make a notation simplification for integrals over the initial or combined functions, useful for calculating moments:
\begin{equation}
    \int_{-\infty}^{\infty} f(x) P_{i,c}(x)dx = \langle f(x) \rangle_{i,c}
\end{equation}
where $i$ or $c$ denotes which function is used in the integration.  We thus denote the moments of our combined distribution as
\begin{equation}
    m_n = \langle(x-\alpha)^n\rangle_c .
\end{equation}
Now we can begin expressing the polydispersity of $P_c(x)$, when we note $\langle(x-\alpha)^n\rangle_c=\langle(\alpha(x-\alpha)^n)\rangle_i+(1-\alpha)(-\alpha)^n$.
\begin{equation}
    \delta=\frac{\sqrt{m_2}}{\mu} =
     \frac{\sqrt{\alpha \langle (x-\alpha)^2 \rangle_i+(1-\alpha)\alpha^2}}{\alpha \langle x \rangle_i+(1-\alpha)*0}
    \label{appendpoly}
\end{equation}
We can establish the following equations to help us in these calculations:
\begin{eqnarray}
    \langle \alpha \rangle_i &=& \alpha \label{appendtrick1}\\
    \langle x \rangle_i &=& 1\\
    \langle (x-1)^2 \rangle_i &=& \delta_i^2 \label{appendtrick2}\\
    \langle (x-1)^3 \rangle_i &=& S_i\delta_i^{2/3} \label{appendtrick3}
\end{eqnarray}
These can be rearranged to give expressions for $\langle x^2 \rangle_i$ and $\langle x^3 \rangle_i$ in terms of $\delta_i, S_i$, and constants.  This then lets us evaluate the averages in Eq.~\ref{appendpoly} leading to
\begin{equation}
    \delta = \sqrt{\frac{\delta_i^2+1}{\alpha}-1}.
\end{equation}
We can then solve for $\alpha$:
\begin{equation}
    \alpha=\frac{\delta_i^2+1}{\delta^2+1} .
    \label{alphafuncddi}
\end{equation}
This is useful as we can later get an expression for the skewness which will not explicitly depend on $\alpha$.

Now we can do the same method for skewness:
\begin{equation}
S_1=\frac{m_3}{m_2^{3/2}}=\frac{\alpha \langle (x-\alpha)^3 \rangle_i-(1-\alpha)\alpha^3}{\left(\alpha \langle (x-\alpha)^2 \rangle_i+(1-\alpha)\alpha^2\right)^{3/2}}.
\end{equation}
Again we can use the results of Eqs.~\ref{appendtrick1} - \ref{appendtrick3} to compute the averages, leading to
\begin{equation}
    S_1=\frac{S_i\delta_i^3+(3-3\alpha)\delta_i^2+2\alpha^{2}-3\alpha+1}{\alpha^{1/2}((\delta_i^2+1)-\alpha)^{3/2}}.
\end{equation}
Equation \ref{alphafuncddi} can be used to substitute for $\alpha$.  The distributions of interest have $S_i=0$, leading to the minimum packing occurring at skewness
\begin{equation}
    S_1=\frac{\delta^{4}\left(1+3\delta_{i}^{2}\right)-\delta^{2}\left(1+3\delta_{i}^{4}\right)+\delta_{i}^{2}\left(1-\delta_{i}^{2}\right)}{\delta^{3}\left(1+\delta_{i}^{2}\right)^{2}}.
\end{equation}
If one considered initial distributions with an arbitrary $S_i$, this leads to an extra term,
\begin{equation}
    S_1' = S_1 + S_i \frac{\delta_i^3(1+\delta^2)^2}{\delta^3 (1 + \delta_i^2)^2}.
\end{equation}


\end{document}